%% file: eprint_prelovsek.tex
\documentclass[10pt]{article}
\usepackage{graphicx}
 \usepackage{amsmath}
\usepackage{amssymb} 
\usepackage{epsfig}
\usepackage{color}
\usepackage{latexsym}
\usepackage{amsmath}
\usepackage{amssymb}
\usepackage{dsfont}
\usepackage{verbatim}

\def\Title#1{\begin{center} {\Large #1 } \end{center}}
\def\Author#1{\begin{center}{ \sc #1} \end{center}}
\def\Address#1{\begin{center}{ \it #1} \end{center}}

\newcommand\pubblock{\rightline{\begin{tabular}{l} Proceedings of the Fifth Annual LHCP\\ \pubnumber\\
         \pubdate  \end{tabular}}}

\newenvironment{Abstract}{\begin{quotation} \begin{center} 
             \large ABSTRACT \end{center}\bigskip 
      \begin{center}\begin{large}}{\end{large}\end{center} \end{quotation}}

\newenvironment{Presented}{\begin{quotation} \begin{center} 
             PRESENTED AT\end{center}\bigskip 
      \begin{center}\begin{large}}{\end{large}\end{center} \end{quotation}}

\def\Acknowledgements{\bigskip  \bigskip \begin{center} \begin{large}
             \bf ACKNOWLEDGEMENTS \end{large}\end{center}}

\input econfmacros.tex

\usepackage{color}

 \newcommand\pubnumber{ }

\newcommand\pubdate{}

\def\affiliation{Faculty of Mathematics and Physics, University of Ljubljana,   Jadranska 19, Slovenia\\
Jozef Stefan Institute, Jamova 39,   Ljubljana, Slovenia\\
 Instit\"ut f\"ur Theoretische Physik, University of Regensburg, Germany  }

\begin{document}

\large
\begin{titlepage}
\pubblock

\vfill
\Title{   Heavy Flavors on The Lattice  }
\vfill

\Author{  Sasa Prelovsek  }
\Address{\affiliation}
\vfill
\begin{Abstract}
The lattice QCD results on the hadron spectrum and weak transitions between hadrons are briefly reviewed. Hadrons containing heavy quarks $c$ or $b$ are considered. The focus is on the recent simulations  and some older results which are particularly relevant in view of the recent experimental discoveries.   

\end{Abstract}
\vfill

\begin{Presented}
The Fifth Annual Conference\\
 on Large Hadron Collider Physics \\
Shanghai Jiao Tong University, Shanghai, China\\ 
May 15-20, 2017
\end{Presented}
\vfill
\end{titlepage}
\def\thefootnote{\fnsymbol{footnote}}

\setcounter{footnote}{0}
%

\normalsize 


\section{Introduction} 
The study of hadron properties requires a non-perturbative method since the strong coupling constant at hadronic energy is not small. Lattice QCD is an ab-initio non-perturbative method based directly on the QCD Lagrangian with parameters $m_{q_i}$ and $g_s$. 
 An expectation value of a desired quantity $C$ is obtained via numerical path integration  $\langle C \rangle \propto \int {\cal D}U {\cal D} q_i {\cal D} \bar q_i  e^{-S_{QCD}}C $ formulated on a discretized and finite Eucledian space-time.     
 
   Here I focus  on the lattice results reported between the summer 2016 and the LHCP  conference held in mid-May 2017.  
   Some older lattice results  are also reported which are particularly relevant in view of the very recent experimental discoveries.   
 
\section{Spectroscopy  and two-hadron scattering  }

\underline{\bf Excited charmonia, charmed and charmed-strange mesons}: The most extensive spectra of the excited charmonia, $D$ and $D_s$ have been calculated Hadron Spectrum Collaboration.
   Several complete quark-antiquark multiplets $nL$ were found in
a recent simulation   with $m_\pi\simeq 240~$MeV: see Figs. 3, 4, and 5 in  \cite{Cheung:2016bym}. Multiplets of hybrid states were also found   and 
some of them carry exotic $J^{PC}$.     Light-quark mass dependence of charmonia in comparison to earlier results at  $m_\pi\simeq 400~$MeV is found to be mild. 
Most of these states, particularly those with $J=3,4$ or exotic $J^{PC}=1^{-+},~ 0^{+-}, ~2^{+-},..$ have yet to be discovered experimentally.   
The main caveat of this calculation is that it disregards  strong decays of resonances and threshold effects, which is remedied for some states in   what follows. \\

\underline{\bf Strongly decaying hadron resonances and hadron-hadron scattering}: The hadronic resonances with masses above  threshold strongly decay 
to a pair of hadrons $H_1H_2$.  In the recent years, the main effort was to simulate $H_1H_2$ scattering channels on the lattice and extract the underlying scattering matrix $T(E)$ as a function of energy. This is possible thanks to the rigorous L\"uschers formalism \cite{Luscher:1991cf} (see \cite{Prelovsek:2011nk} for introduction on the topic). The scattering matrix $T(E)$ renders the $H_1H_2$ cross-section $\sigma(E)\propto |T(E)|^2$, which in principle allows lattice QCD determination of  the resonance masses and decay widths  from the peaks in the cross-section.  \\

\underline{\bf Charmed resonances   from $D\pi-D\eta-D_s\bar K$ coupled channel scattering}: 
Hadron Spectrum Collaboration extracted $3\times 3$ scattering matrix for three coupled  channels at $m_\pi\simeq400$ MeV and searched for poles in the scattering matrix \cite{Moir:2016srx}. The resonance pole was found in $d-$wave scattering, which closely resembles  experimental $D_2$ with $J^P=2^+$. The scalar $D_0^*$ state was found as a bound-state pole on the real-axes   almost on $D\pi$ threshold due to the heavy $m_{\pi}$ employed in the simulation \cite{Moir:2016srx}. Experimentally scalar $D_0^*(2400)$ meson is a wide resonance above $D\pi$ threshold;  it emerged as a wide resonance in a less detailed 
  lattice simulation of $D\pi$ scattering   in the one-channel approximation  at $m_\pi\simeq 266$ MeV \cite{Mohler:2012na}. \\
  
  \underline{ \bf $Z_c(3900)$ from coupled channel $J/\psi\pi-D\bar D^*-\eta_c\rho$ scattering}: In order to search for the exotic $Z_c(3900)$ state with flavor $\bar cc \bar du$, 
  the HALQCD collaboration extracted $3\times 3$ scattering matrix for three coupled channels with $J^P=1^+$ \cite{Ikeda:2016zwx}. Less rigorous HALQCD approach was applied, which has not been verified on conventional resonances. The resulting differential ratios as a function of $J/\psi\pi$ and $D\bar D^*$ invariant masses indeed show a peak around $3.9~$GeV,  resembling experimental peak. If the coupling between  $J/\psi\pi-D\bar D^*$ channels is set to zero by hand, the peak disappears. The HALQCD results therefore indicate that $Z_c(3900)$ peak is possibly a coupled channel effect rather than a genuine resonance. \\

\underline{\bf Search for $X(5568)$  in $B_s\pi^+$ scattering}: 
 The X(5568) state with exotic content $\bar b s\bar du$ was recently  claimed by D0 collaboration \cite{D0:2016mwd}.
 If this state with $J^P=0^+$ exists, it can strongly decay only to $B_s \pi^+$
and lies significantly below all other thresholds, which makes a lattice search
for X(5568) cleaner and simpler than for other exotic candidates.  The
simulation of $B_s\pi^+$ scattering  did not find  X(5568) \cite{Lang:2016jpk},
in  agreement with the  LHCb~\cite{Aaij:2016iev} and CMS~\cite{CMS:2016fvl}
results.  \\

\underline{\bf Doubly bottom $BB^*$ and $B_sB^*$ bound states}: Several lattice simulations provide a growing evidence for a strongly-stable state   with flavor $\bar b\bar bud$, $J^P=1^+$, $I=0$  and  mass $m<m_B+m_{B^*}$. The simulation \cite{Bicudo:2015vta,Bicudo:2016ooe} was based on the static $b$-quark, while \cite{Francis:2016hui} employed NRQCD for $b$-quark. Such a state, if bound by  $m_B+m_{B^*}-m=189\pm 10$ MeV \cite{Francis:2016hui}, decays only weakly to $ud\bar b\bar b \to B^+\bar D^0,\ J/\psi \;B^+K^0$. The indication for a strongly-state strange partner $B_sB^*$  with $m<m_{B_s}+m_{B^*}$ was also found \cite{Francis:2016hui}, with expected weak decays to $B^+D_s^-,\ J/\psi B_sK^0, \ B_s\bar D^0, \ J/\psi B^0\phi$. \\


\underline{\bf Excited $\Omega_c^*$}: The extensive excited $\Omega_c^*$ spectrum with $css$ valence content was predicted on lattice in 2013 \cite{Padmanath:2013bla,Padmanath:2017lng}, disregarding  their strong decays. Five states were found in the energy region $3.0-3.2$ GeV, in agreement with  LHCb discovery this year. The lattice  calculation predicts their quantum numbers: two carry $J^P=1/2^-$, two carry $3/2^-$ and one $5/2^-$. \\

\underline{\bf  Charmonium scalar resonance $\chi_{c0}(2P)$}: This year Belle reported on an alternative candidate for a scalar resonance $\chi_{c0}(2P)$ with $m=3862{+26\ +40\atop -32\ -13}$ MeV and $\Gamma=201{+154\ +88\atop -67\ -82}$ MeV which favors $J^{PC}=0^{++}$ \cite{Chilikin:2017evr}.  A slightly heavier and narrower resonance with $m=3966\pm 20$ MeV and $m=67\pm 18$ MeV emerged from exploratory simulation of $D\bar D$ scattering in s-wave \cite{Lang:2015sba} and  a Breit-Wigner-type fit  in the resonance region (scenario (i) from \cite{Lang:2015sba}).  The uncertainty on the width is large in both cases, and  more   work is needed to understand the physics in this channel. \\

\underline{\bf Hadro-quarkonium}: A system of a quarkonium $\bar QQ$ and a light hadron (meson or baryon) was studied \cite{Alberti:2016dru}. The modifications on $\bar Q(r)Q(0)$ binding potential $V(r)$ was found to be at most  few MeV. Consequently also $\bar QQ$ binding energies modify by at most few MeV in presence of a light hadron.    

\section{Weak transitions}

 \underline{\bf Exclusive $b\to c l\bar \nu$ decays:} Most recent calculations investigate this transition in view of the tension between 
$V_{cb}$ obtained from exclusive and inclusive decays.
 
The new preliminary results for $B\to D^*$ at zero $D^*$ recoil have been obtained by HPQCD,  based on relativistic HISQ charm quark and NRQCD bottom quark \cite{Harrison:2016gup}. The resulting  form factor $h_{A_1}(1)$ leads to the preliminary value of $V_{cb}=41.5(17) \cdot 10^{-3}$ \cite{Harrison:2016gup} that is closer to inclusive  $V_{cb}$ than previous  exclusive $V_{cb}$  from FNAL/MILC  \cite{Bailey:2014tva}. The updated final results from this HPQCD 
simulation are expected soon.   

The preliminary status of the first on-going $B\to D^*$ simulation at non-zero $D^*$ recoil was presented at the  Lattice 2017  (held after LHCP 2017)  by Alejandro Vaquero on behalf of the Fermilab/MILC collaborations. This will be very valuable to verify the Standard Model prediction for $R(D^*)=Br(B\to D^*\tau \bar \nu_\tau)/Br(B\to D^*\mu \bar \nu_\mu)$ \cite{Fajfer:2012vx}, which shows intriguing tension with the experiment. 

\vspace{0.1cm}
  
 The $f_0$ and $f_+$ form factors for  $B_s\to D_s$ were extracted in the whole $q^2$ region  by HPQCD  \cite{Monahan:2017uby}.   Relativistic HISQ c-quark and NRQCD b-quark were employed. These  are valuable also since $f_0^{B_s\to D_s}(m_\pi^2)/f_0^{B\to D}(m_K^2)$ allows determination of   fragmentation-function ratio $f_s/f_d$ needed for LHC measurement of  $B(B_s\to \mu^+\mu^-)$    due to the normalization modes.  The HPQCD determined $f_s/f_d$  using   experimental $B(\bar B_s\to D_s^+\pi^-)/B(\bar B^0\to D^+K^-)$  and the above form-factor ratio   via the factorization hypothesis as proposed in \cite{Fleischer:2010ay}. 
 
 \vspace{0.1cm}
 
 Two methods of treating $b$ quarks were shown to give consistent results for $B_c\to \eta_c$ and $B_c\to J/\psi$ form factors in HPQCD study \cite{Colquhoun:2016osw}.   First method considers relativistic HISQ   $b$-quark on a range of masses $m_Q\leq m_b$, while NRQCD is used in the second method: agreement at $m_Q=m_b$ gives credibility in the results.  \\
 
    \underline{\bf Towards inclusive semileptonic decays}: The formalism and first exploratory lattice results  for inclusive semileptonic decays  $B\to X_c l\bar \nu$ were presented by S. Hashimoto \cite{Hashimoto:2017wqo}.  The squared amplitude $|{\cal M}|^2=|V_{cb}|^2 G_F^2 M_B ~ l^{\mu\nu} ~W_{\mu\nu}$ based on $H=V_{cb} \tfrac{G_f}{\sqrt{2}} ~ \bar l \gamma_\mu (1-\gamma_5)\nu J^\mu$ ($J^\mu\equiv \bar c \gamma^\mu (1-\gamma_5)b=V^\mu-A^\mu$)    contains trivial leptonic part $l^{\mu\nu}$, while the hadronic part   
 $$W_{\mu\nu}=\sum_X(2\pi)^3 \delta^4(p_B-q-p_x)\frac{1}{2M_B}~\langle B(p_B)|J_\mu^\dagger(0)|X\rangle ~\langle X| J_\nu(0)|B(p_B)\rangle$$ 
  contains a sum over all  on-shell final states $X_c=D,D^*,...$.  Instead of summing those explicitly, one can obtain    the sum directly by considering forward scattering matrix element 
  $$T^{\mu\nu}=i \int d^4x ~e^{-iqx} \frac{1}{2M_B} \langle B|T\{J_\mu^\dagger(x)J_\nu(0)\}|B\rangle$$
  which is related to desired $W$ via the optical theorem  $2\; \mathrm{Im} M(B\to B)=\sum_X \int d\Pi_X |M(B\to X)|^2$ (see for example Section 18.5 of Peskin) as
  $W=-\frac{1}{\pi}~ \mathrm{Im} T$. The $T$ can be  computed on the lattice from the four-point function \cite{Hashimoto:2017wqo}
  $$T^{JJ}_{\mu\nu}(\omega,\vec q)\propto \int_0^\infty dt ~e^{\omega t} \int d\vec x~e^{i\vec q \vec x} ~\langle 0|B(\vec p_B\!\!=\!\!\vec 0,t_{snk}) ~J^\dagger_\mu(\vec x,t_2)~J_\nu(0,t_1) ~B^\dagger(\vec 0,t_{src})|0\rangle$$
  where $\omega$ denotes the energy of the final state $X$,  $J$ contains $V$ or $A$ parts and various normalizations factors have been omitted for simplicity.  
  
  The exploratory numerical simulation has been done on JLQCD configurations for the spectator $s$ quark $B_s\to X_c l\bar \nu$ at zero recoil $\vec q=\vec p_B-\vec p_X=\vec 0$ and for $b$-quark mass smaller than physical.   The resulting  $T_{\mu\nu}(\omega,\vec q=\vec 0)$ as a function of $\omega$ is shown in Fig. 10 of \cite{Hashimoto:2017wqo}. The $V_0V_0$ part is found to be dominated by $D_s$ pole (dashed lines) and represents another way to extract of HQET form factor $h_+(1)$ via 
  $T_{00}^{VV}=\frac{|h_+(1)|^2}{M_{D_s}-\omega}$. The   $T_{11}^{AA}(\omega, \vec 0)= \frac{|h_{A_1}(1)|^2}{M_{D_s^*}-\omega}$  is dominated by $D_s^*$ pole. The HQET form factors determined in this way are consistent with direct calculation but have currently larger errors.    The $V_1V_1$ and $A_0A_0$  contain contribution from $X_c$ final states with other quantum numbers. \\

   \underline{\bf New  exclusive $q\to q^\prime l\bar \nu$ and $q\to q^\prime l^+l^-$ baryon decays}: Six $\Lambda_c\to \Lambda l\bar \nu$ form factors were determined \cite{Meinel:2016dqj} for the first time in view of the  BESIII 2015 measurement of this decay rate.  Taking $V_{cs}$ from global CKM fit, these form factors lead to the rate that is consistent, and twice more precise as BESIII rate. This gives confidence in lattice treatment of this and analogous electroweak baryon transitions.  Alternatively, the form factors lead to the value of $V_{cs}$, which is currently less precise, but consistent with the one from $D_s\to l\bar \nu$. 

The $b\to sl^+l^-$ transition and their ratios (relevant for the lepton flavor violation) have attracted large attention recently in view of few tensions between Standard model predictions and experiment. 
On this front, there is only one new lattice result available since summer 2016. This is a report on the ongoing simulation of $\Lambda_b\to \Lambda^*(1520) l^+l^-$ form factors \cite{Meinel:2016cxo}. The   $\Lambda_b\to \Lambda^* l^+l^-$  has namely certain advantages with respect to the more standard $\Lambda_b\to \Lambda l^+l^-$ decay where   $\Lambda$ is neutral and long-lived, which is not favorable for the accurate experimental study.  The unstable $\Lambda^*$ resonances, one the other hand,  immediately decay into charged particles and produce tracks that originate from $b$-decay vertex, which motivates exploring decays  also through unstable $\Lambda^*$ in experiment. 
 
 \section{Conclusions}
 
    Lattice QCD is a reliable ab-initio non-perturbative method that is based directly on the fundamental theory of strong interactions - QCD.

Spectra of strongly stable hadrons ($B$, $D$, ...) are well under control  and in agreement with experiment. Variety of exclusive electro-weak transitions between them are being studied with increasing precision. The formalism and the first exploratory lattice results for  the inclusive weak transitions $B\to X_c l\bar \nu_l$ have been presented during the last year. 

Experiments provided lots of interesting and puzzling hadrons in the recent years. The five   excited $\Omega_c^*\simeq css$ states, discovered by LHCb this year, have been predicted by lattice QCD around 2013.   There is a growing evidence for a strongly-stable state   with flavor $\bar b\bar bud$, $J^P=1^+$, $I=0$  and  mass $m<m_B+m_{B^*}$ from the lattice. Most of the experimentally discovered   exotic hadrons are actually strongly decaying resonances that decay to a single   or to multiple-channels. They have to be inferred from the one-channel   or multiple-channel scattering matrix.  It is encouraging that  scattering matrices for single-channel scattering have been  reliably extracted for certain channels in the recent years. Scattering matrices for two-channel and three-channel scattering have also been extracted in some cases. One  study along these lines  indicates that the  experimental $Z_c^+(3900)$ peak  arises due to the large coupling between the $D\bar D^*$ and $J/\psi\pi^+$ channels.   Lots of exciting and challenging problems along these lines still remain to be attacked.

\Acknowledgements
I would like to thank  G. Bali, P. Bicudo, S. Hashimoto, J. Harrison, L. Leskovec, A. Lytle, M. Padmanath and M. Wingate for discussions or for sending their  results. This work is supported by  the Slovenian Research Agency ARRS (research core funding No. P1- 0035).


\end{document}

%% file: econfmacros.tex
\def\beq{\begin{equation}}
\def\eeq#1{\label{#1}\end{equation}}
\def\eeqn{\end{equation}}

\def\beqa{\begin{eqnarray}}
\def\eeqa#1{\label{#1}\end{eqnarray}}
\def\eeqan{\end{eqnarray}}

\let\bar=\overbar

\def\Dslash{\not{\hbox{\kern-4pt $D$}}}
\def\dslash{\not{\hbox{\kern-2pt $\del$}}}

\def\msb{{\bar{\ssstyle M \kern -1pt S}}}

